\documentclass[prl,twocolumn,showpacs,superscriptaddress,preprintnumbers,amssymb]{revtex4}
\usepackage{graphicx}% Include figure files
\usepackage{graphics}% Include figure files in another version
\usepackage{bm}% bold math
\newcommand{\beq}{\begin{equation}}
\newcommand{\eeq}{\end{equation}}
\newcommand{\beqn}{\begin{eqnarray}}
\newcommand{\eeqn}{\end{eqnarray}}

\newcommand{\ua}{\uparrow}

\newcommand{\figref}[1]{Fig.\,\ref{#1}}
\newcommand{\eqnref}[1]{Eq.\,(\ref{#1})}

\begin{document}

\title{Wave Function and Strange Correlator of Short Range Entangled states}

\author{Yi-Zhuang You}

\author{Zhen Bi}

\author{Alex Rasmussen}

\author{Kevin Slagle}

\author{Cenke Xu}

\affiliation{Department of physics, University of California,
Santa Barbara, CA 93106, USA}

\begin{abstract}

We demonstrate the following conclusion: If $|\Psi\rangle$ is a
$1d$ or $2d$ nontrivial short range entangled state, and $|\Omega
\rangle$ is a trivial disordered state defined on the same Hilbert
space, then the following quantity (so called strange correlator)
$C(r, r^\prime) = \frac{\langle \Omega|\phi(r) \phi(r^\prime) |
\Psi\rangle}{\langle \Omega| \Psi\rangle}$ either saturates to a
constant or decays as a power-law in the limit $|r - r^\prime|
\rightarrow +\infty$, even though both $| \Omega\rangle$ and $|
\Psi\rangle$ are quantum disordered states with short-range
correlation. $\phi(r)$ is some local operator in the Hilbert
space. This result is obtained based on both field theory
analysis, and also an explicit computation of $C(r, r^\prime)$ for
four different examples: $1d$ Haldane phase of spin-1 chain, $2d$
quantum spin Hall insulator with a strong Rashba spin-orbit
coupling, $2d$ spin-2 AKLT state on the square lattice, and the
$2d$ bosonic symmetry protected topological phase with $Z_2$
symmetry. This result can be used as a diagnosis for short range
entangled states in $1d$ and $2d$.

\end{abstract}

\date{\today}

\maketitle

A short range entangled (SRE) state is a ground state of a quantum
many-body system that does not have ground state degeneracy or
bulk topological entanglement entropy. But a SRE state (for
example the integer quantum Hall state) can still have protected
stable gapless edge states. Thus it appears that the bulk of all
the SRE states are identically trivial, and a nontrivial SRE state
is usually characterized by its edge states. In this paper we
propose a diagnosis to determine whether a given many-body wave
function defined on a lattice without boundary is a nontrivial SRE
state or a trivial one. This diagnosis is based on the following
quantity called ``strange correlator" \footnote{In the
thermodynamic limit, both numerator and denominator of the strange
correlator approach zero, while their ratio remains a finite
constant. All the calculations in this paper were based on finite
system size first, the thermodynamic limit is taken {\it after}
taking the ratio.}: \beqn C(r, r^\prime) = \frac{\langle
\Omega|\phi(r) \phi(r^\prime) | \Psi \rangle}{\langle \Omega|
\Psi\rangle}. \label{sc}\eeqn Here $|\Psi\rangle$ is the wave
function that needs diagnosis, $|\Omega\rangle$ is a direct
product trivial disordered state defined on the same Hilbert
space. Our conclusion is that if $|\Psi\rangle$ is a nontrivial
SRE state in one or two spatial dimensions, then for some local
operator $\phi(r)$, $C(r, r^\prime)$ will either saturate to a
\emph{constant} or decay as a \emph{power-law} in the limit $|r -
r^\prime| \rightarrow +\infty$, even though both $| \Omega
\rangle$ and $| \Psi\rangle$ are disordered states with
short-range correlation.

Another possible way of diagnosing a SRE wave function is through
its entanglement spectrum~\cite{haldaneentangle}. If a SRE state
has nontrivial edge states, an analogue of its edge states should
also exist in its entanglement spectrum~\cite{qientangle}.
However, many SRE states are protected by certain symmetry, some
SRE states are protected by lattice symmetries (for example the
spin-2 AKLT state on the square lattice requires translation
symmetry). If the cut we make to compute the entanglement spectrum
breaks such lattice symmetry, then the entanglement spectrum would
be trivial, even if the original state is a nontrivial SRE state.
By contrast, the strange correlator in \eqnref{sc} is defined on a
lattice without edge, thus it already preserves all the symmetries
of the system, including all the lattice symmetries. Thus the
strange correlator can reliably diagnose SRE states protected by
lattice symmetries as well.

\begin{figure} %[htbp]
\begin{center}
\includegraphics[width=220pt]{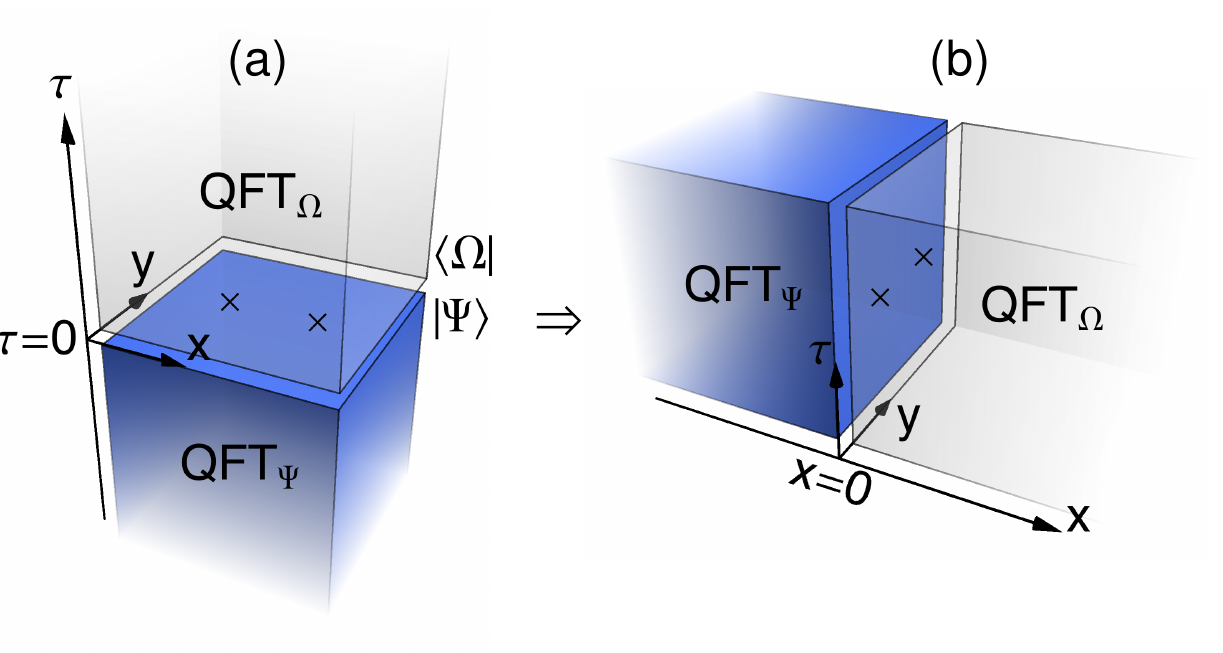}
\caption{(color online). (a) $|\Psi\rangle$ and $\langle\Omega|$
are given by infinite time evolution of their quantum field
theories (QFT) from below and above respectively. The strange
correlator can be viewed as the correlator at the $\tau=0$
interface. (b) Under the Lorentz rotation, the two QFT's are
separated by the $x=0$ interface, and the strange correlator can
be interpreted as the correlation function on this spatial
interface.} \label{fig: lorentz}
\end{center}
\end{figure}

The strange correlator can be roughly understood as follows: $|
\Psi\rangle$ can be viewed as a generic initial state evolved with
a constant nontrivial SRE Hamiltonian from $\tau = - \infty$ to
$\tau = 0$; $\langle \Omega|$ is a state evolved from $\tau = +
\infty$ to $\tau = 0$ with a trivial Hamiltonian, thus the strange
correlator can be viewed as a ``correlation function" at a
temporal domain wall of the QFT's at $\tau = 0$, see \figref{fig:
lorentz}(a). If there is an approximate Lorentz invariant
description of the system, a space-time rotation can transform
\eqnref{sc} to a space-time correlation at a spatial interface
between nontrivial and trivial SRE systems, see \figref{fig:
lorentz}(b). And for one and two spatial dimensions, a spatial
interface between trivial and nontrivial SRE states should have
either long range or power-law correlation between certain local
operators, which after Lorentz rotation will lead to the
conclusion of this paper. A similar observation of Lorentz
rotation was used to derive the bulk wave function of topological
superconductors~\cite{shankar2011}.

For bosonic SRE states that are protected by certain symmetry (so
called symmetry protected topological (SPT)
states~\cite{wenspt,wenspt2}), the argument above can be
demonstrated more explicitly. In Ref.\,\cite{xuclass}, it was
demonstrated that a large class of $1d$ and $2d$ bosonic SPT
states can be described by the following two nonlinear Sigma model
(NLSM) field theories: \beqn \mathcal{S}_{1d} = \int dx d\tau
\frac{1}{g} (\partial_\mu \vec{n})^2 + \frac{i 2\pi}{8\pi}
\epsilon_{abc}\epsilon_{\mu\nu} n^a \partial_\mu n^b
\partial_\nu n^c, \label{o3theta}\eeqn \beqn
\mathcal{S}_{2d} = \int d^2x d\tau \frac{1}{g} (\partial_\mu
\vec{n})^2 + \frac{i 2\pi}{12 \pi^2}
\epsilon_{abcd}\epsilon_{\mu\nu\rho} n^a
\partial_\mu n^b
\partial_\nu n^c \partial_\rho n^d. \label{o4theta} \eeqn Here
$\vec{n}(x)$ is an O(3) or O(4) vector order parameter with unit
length constraint: $(\vec{n})^2 = 1$. Different SPT phases are
distinguished from each other based on different implementations
of the symmetry group on the vector order parameter $\vec{n}$. In
both $1d$ and $2d$, ground state wave functions of SPT phases can
be derived based on \eqnref{o3theta} and \eqnref{o4theta} (see
Ref.\,\cite{xusenthil}): \beq |\Psi\rangle_{d} \sim \int  \
D\vec{n}(x) \ \exp^{ - \int_{S^d} d^dx \ \frac{1}{G} (\nabla
\vec{n})^2 + \mathrm{WZW}_d[\vec{n}]} | \vec{n}(x) \rangle,
\label{wf} \eeq where $S^d$ is the compactified real space
manifold, and $\mathrm{WZW}_d[\vec{n}]$ is a real space
Wess-Zumino-Witten term: \beqn \mathrm{WZW}_1[\vec{n}] &=&
\int_0^1 du \ \frac{i 2\pi}{8\pi} \epsilon_{\mu\nu} \epsilon_{ab}
n^a \partial_\mu n^b \partial_\nu n^c, \ \mu, \nu = x, u \cr\cr
\mathrm{WZW}_2[\vec{n}] &=& \int_0^1 du \ \frac{i 2\pi}{12 \pi^2}
\epsilon_{abcd}\epsilon_{\mu\nu\rho} n^a
\partial_\mu n^b \partial_\nu n^c \partial_\rho n^d, \cr \cr \mu, \nu, \rho &=& x, y, u.
\eeqn In contrast, the trivial state wave function is a
superposition of all configurations of $|\vec{n}(x)\rangle$
without a WZW term. Based on the wave functions in \eqnref{wf},
the strange correlator of order parameter $\vec{n}(x)$ reads \beqn
C(r, r^\prime) = \frac{ \int D\vec{n}(x) \ n^a(r) n^b(r^\prime)
e^{ - \int_{S^d} d^dx \ \frac{1}{G} (\nabla \vec{n})^2 +
\mathrm{WZW}_d[\vec{n}]} }{\int D\vec{n}(x) \ e^{ - \int_{S^d}
d^dx \ \frac{1}{G} (\nabla \vec{n})^2 + \mathrm{WZW}_d[\vec{n}]}}.
\eeqn Mathematically, this strange correlator can be viewed as an
ordinary space-time correlation function of a $(d-1)+1$
dimensional field theory with a WZW term, as long as we view one
of the spatial coordinate as the time direction. When $d = 1$,
this strange correlator is effectively a spin-spin correlation of
one isolated free spin-1/2, and the correlation is always long
range. When $d = 2$, this strange correlator is effectively a
space-time correlation function of a $(1+1)d$ O(4) NLSM with a WZW
term, and when this model has a full SO(4) symmetry, this theory
is a SU(2)$_1$ conformal field theory with power-law
correlation~\cite{witten,witten1984}; when the symmetry of the
system is a subgroup of SO(4), as long as the residual symmetry
prohibits any linear Zeeman coupling to order parameter $\vec{n}$,
this $(1+1)d$ system either remains gapless, or spontaneously
breaks the symmetry and develop long range order. Thus the strange
correlator is either long range or decays with a power-law.

The two arguments above both rely on a certain continuum limit
description of the SRE state. However, for a fully gapped system,
when the gap is comparable with the ultraviolet energy scale of
the system, a continuum limit description may not be appropriate.
In the rest of the paper, we will compute the strange correlator
for several examples of SRE states {\it far from} the continuum
limit, $i.e.$ the gap of the SRE states is comparable with UV
cut-off. We will see that in some examples, the strange correlator
is indeed different from the physical edge of the SRE state, but
our qualitative conclusion is still valid.

The first example we study is the AKLT
state~\cite{affleck1987,affleck1988} of the Haldane phase of
spin-1 chain. In the $S^z$ basis, the AKLT wave function is a
``dilute" N\'{e}el state, namely it is an equal weight
superposition of all the $S^z$ configurations with an alternate
distribution of $|+\rangle = | S^z = +1 \rangle$ and $|-\rangle =
| S^z = -1 \rangle$, sandwiched with arbitrary numbers of
$|0\rangle = | S^z = 0 \rangle$~\cite{string}: \beqn | \Psi
\rangle = \sum \frac{1}{N} |+ 0 \cdots 0 - 0\cdots 0 +
\cdots\rangle \eeqn We choose the trivial state to be $| \Omega
\rangle = | 0 0 0 \cdots \rangle$. Straightforward calculation
leads to the following answer of the strange correlator: \beqn
C(r, r^\prime) = \frac{ \langle \Omega| S^+_r \ S^-_{r^\prime}
|\Psi \rangle } { \langle \Omega |\Psi \rangle }  = 2, \eeqn which
is the expected long range correlation.

The second example we study is the two dimensional quantum spin
Hall (QSH) insulator with a Rashba spin orbit coupling. We will
use the same notation as Ref.\,\cite{kane2005b}. The QSH insulator
Hamiltonian reads \beqn H = t \sum_{\langle i,j \rangle}
c_{i}^\dagger c_j + i\lambda_{SO}\sum_{\langle \langle i,j \rangle
\rangle} \nu_{ij} c^\dagger_i s^z c_j \cr\cr + \lambda_R
\sum_{\langle i,j \rangle} c^\dagger_i (\mathbf{s} \times
\hat{\mathbf{d}}_{ij})_z c_j + \lambda_v \sum_i \xi_i c^\dagger_i
c_i. \label{km}\eeqn $\lambda_{SO}$ is the spin-orbit coupling
that leads to the QSH topological band structure, $\lambda_R$ is
the Rashba coupling that breaks the conservation of $s^z$, and
$\lambda_v$ is a staggered potential that leads to charge density
wave. The electron operator $c_{i}$ carries spin and sublattice
indices, thus the strange correlator $C(r, r^\prime)$ is a $4
\times 4$ matrix. For QSH state $|\Psi\rangle$, we choose $
\lambda_{SO} = t $, $\lambda_R = 0.5t$, $\lambda_v = 0$; trivial
state $|\Omega\rangle$ is chosen to be a strong CDW state with $
\lambda_{SO} = t $, $\lambda_R = 0.5t$, $\lambda_v = 10t$. These
two states are far from the continuum limit, namely the gap is
comparable with the UV cut-off.

\begin{figure} %[htbp]
\begin{center}
\includegraphics[width=210pt]{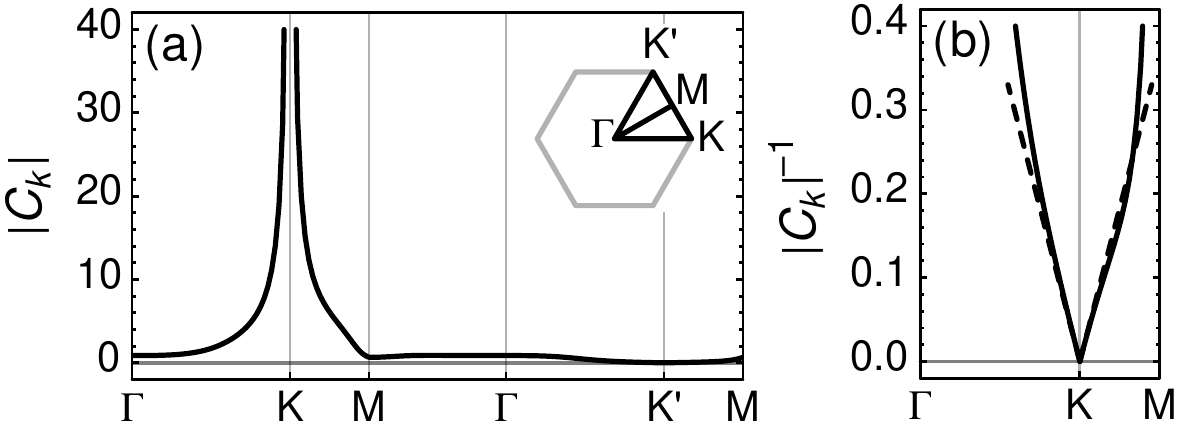}
\caption{(a) The amplitude of strange correlator in the momentum
space. The inset shows the Brillouin zone and the high symmetry
points. (b) $|C_k|^{-1}$ exhibits nice linearity around the $K$
point, establishing the $1/|k|$ divergence in $|C_k|$.}
\label{fig: QSH}
\end{center}
\end{figure}

\figref{fig: QSH}(a) shows the amplitude of strange correlator
$|C_k| = |\langle \Omega | c^\dagger_{A, \ua, k} c_{B, \ua, k}|
\Psi \rangle/ \langle \Omega |\Psi \rangle |$ plotted in the
momentum space. There is one clear singularity at the corner of
the Brillouin zone, which diverges as $\sim 1/|k|$. This implies
that in the real space the strange correlator decays as $|C(r,
r^\prime)| \sim 1/|\vec{r} - \vec{r}^\prime|$, which is consistent
with the result obtained from Lorentz transformation, despite the
large bulk gap.

The third example we will study is the spin-2 AKLT state on the
square lattice,\cite{affleck1988,kennedy1988} which is a SPT state
protected by the on-site $\mathbb{Z}_2\times\mathbb{Z}_2$ and the
lattice translation symmetry,\cite{Liu:2011ek} whose wave function
has a tensor product state (TPS) representation\cite{Wei:2012uo,
Miyake:2011th}
\begin{equation}
|\Psi\rangle = \sum_{\{m_i\}}\mathrm{tTr}(\otimes_i T^{m_i})
|\{m_i\}\rangle.
\end{equation}
Here $m_i=0,\pm1,\pm2$ labels the $S^z$ quantum number of the
spin-2 object on site $i$, and $|\{m_i\}\rangle$ is the state for
the configuration $\{m_i\}$ over the lattice. $\mathrm{tTr}$
traces out the internal legs in the tensor network shown in
\figref{fig: AKLT}(a), in which the vertex tensor is given by
\begin{equation}
T^m_{s_1s_2s_3s_4}= \left\{
    \begin{array}{ll}
        4s_1s_2 & :-s_1-s_2+s_3+s_4=m,\\
        0 & :\text{otherwise},
    \end{array}
\right.
\end{equation}
with $s_j=\pm1/2$ labeling the spin-1/2 internal degrees of
freedom. While the trivial state $|\Omega\rangle = |\{\forall i:
m_i=0\}\rangle$ is chosen to be the direct product state of
$S^z=0$ on every site. We look into the strange correlator
\begin{equation}\label{eq: C of 2dAKLT}
C(r,r')=\frac{\langle\Omega|S_r^+S_{r'}^-
|\Psi\rangle}{\langle\Omega|\Psi\rangle}=\frac{\mathrm{tTr}(T^0\cdots
T^1(r)T^{-1}(r')\cdots)}{\mathrm{tTr}(T^0\cdots)},
\end{equation}
which can be expressed as a ratio between two tensor networks: the
denominator is a uniform network of the tensor $T^0$ on each site,
and the numerator is the same network except for impurity tensors
$T^{\pm1}$ on site $r$ and $r'$ respectively.

%Both networks are defined on the square lattice.

\begin{figure} %[htbp]
\begin{center}
\includegraphics[width=190pt]{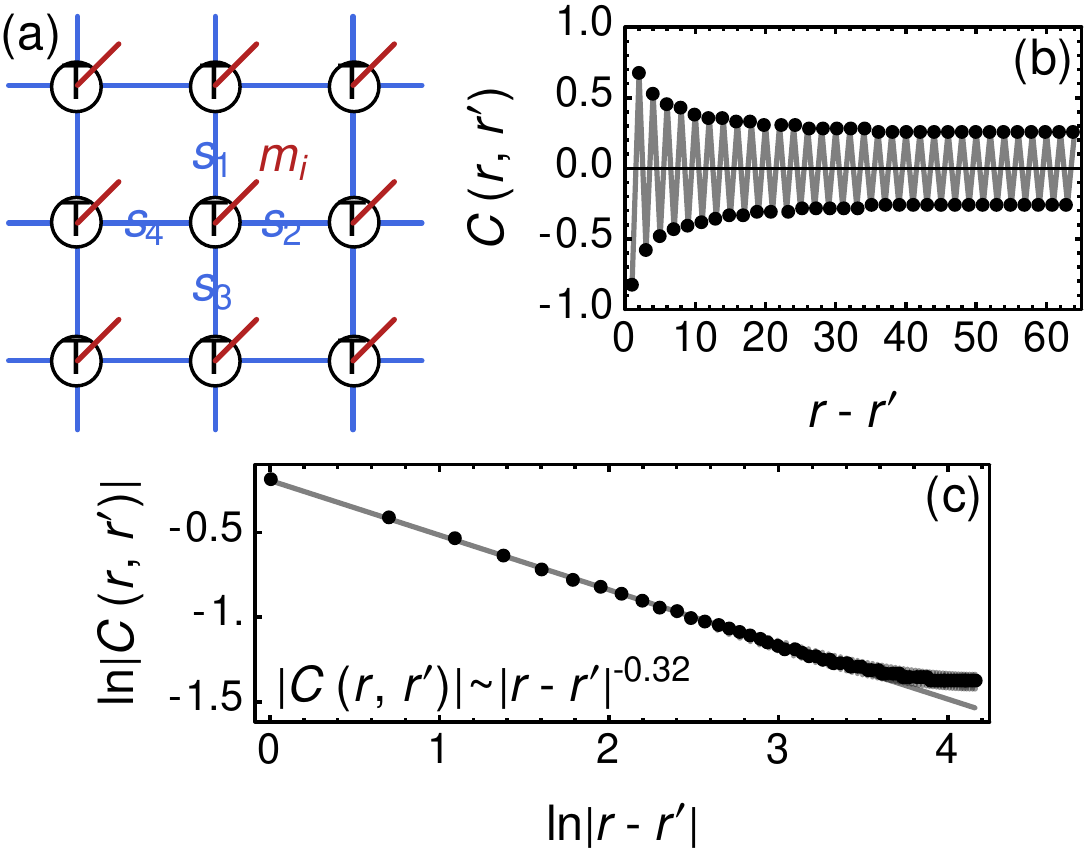}
\caption{(color online). (a) Tensor network representation of the
$2d$ AKLT state. The red (blue) legs represent the physical
(internal) degrees of freedom. (b) Strange correlator of the $2d$
AKLT state measured along the horizontal direction. (c) The
amplitude follows a power-law behavior in the log-log plot. The
final deviation is due to the finite-size effect.} \label{fig:
AKLT}
\end{center}
\end{figure}

The evaluation of the tensor trace in \eqnref{eq: C of 2dAKLT}
over the $2d$ lattice can be reformulated as an (1+1) dimensional
quantum mechanics problem in terms of the transfer matrix for each
row, which can then be studied by the density matrix
renormalization group (DMRG)
method.\cite{White:1992ys,Schollwock:2011fy} The calculation is
performed on an $128\times\infty$ lattice with periodic boundary
condition along both directions. We found that the strange
correlator decays with oscillation (as in \figref{fig: AKLT}(b)),
and its amplitude follows a power-law behavior
$|C(r,r')|\sim|r-r'|^{-\eta}$ with the exponent $\eta \simeq
0.32$, see \figref{fig: AKLT}(c), which is consistent with our
previous field theory argument.

The last example we will study is the two dimensional bosonic SPT
phase with $Z_2$ symmetry which was first studied in
Ref.\,\cite{levingu}. The ground state wave function of this SPT
phase is \beqn \label{eq: LG state}|\Psi\rangle =
\sum_{\{\sigma_i\}} (-1)^{N_d}
\exp\Big(-\frac{\beta}{2}\sum_{\langle i,j\rangle} \sigma_i
\sigma_j \Big) |\{ \sigma_i \}\rangle , \eeqn which is a
superposition of all the configurations of the Ising degree of
freedom $|\{ \sigma_i \}\rangle$ with a factor $(-1)$ associated
with each closed Ising domain wall (with $N_d$ being the number of
domain wall loops). The trivial state $|\Omega\rangle$ is simply
an Ising paramagnet, whose wave function is similar to \eqnref{eq:
LG state} but without the domain wall sign structure $(-1)^{N_d}$.
Compared with Ref.\,\cite{levingu}, we have added a factor
$\exp(-\beta/2\sum_{\langle i,j\rangle}\sigma_i \sigma_j)$ to each
Ising configuration to adjust the spin correlation length.

%The state studied in Ref.\,\cite{levingu} corresponds to the
%``fixed point" $\beta = 0$.

\begin{figure} %[htbp]
\begin{center}
\includegraphics[width=190pt]{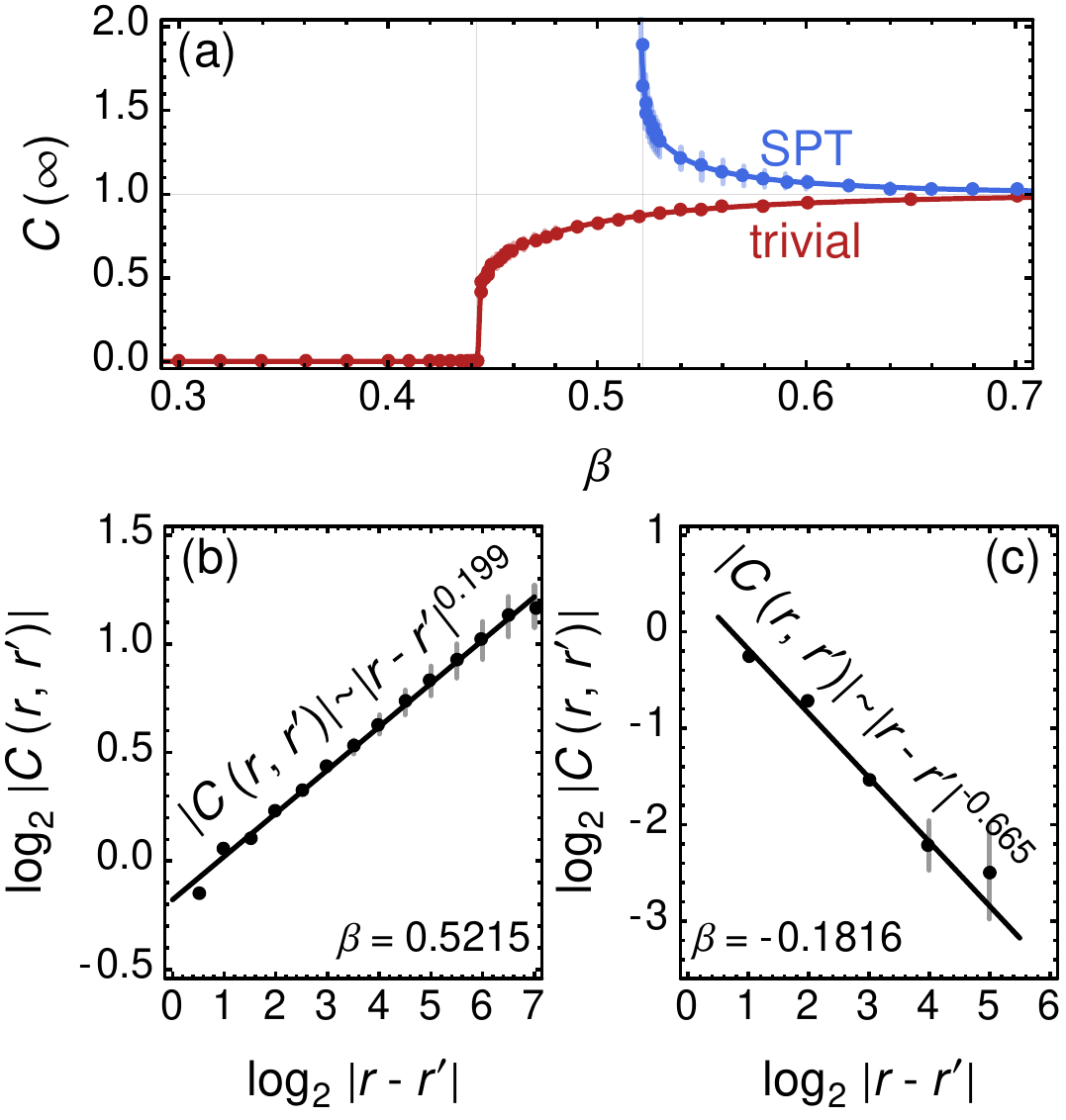}
\caption{(color online). (a) The strange correlator of the SPT
state (in blue) at infinite distance $|r-r'|\to\infty$, in
comparison with that of the trivial state (in red). The SPT
strange correlator follows the power-law behavior (b) at the
critical point and (c) in the dense loop phase.} \label{fig: LG}
\end{center}
\end{figure}

The strange correlator of the $Z_2$ bosonic SPT phase can be
viewed as a correlation function of a ``classical statistical
mechanics model": \beqn C(r, r^\prime) = \frac{ \sum_{\{
\sigma_i\}} \sigma_{r} \sigma_{r^\prime} (-1)^{N_d} e^{ - \beta
\sum_{\langle i,j\rangle} \sigma_i \sigma_j } }{ \sum_{\{ \sigma_i
\}} (-1)^{N_d} e^{ - \beta \sum_{\langle i,j\rangle} \sigma_i
\sigma_j } }. \label{tising} \eeqn Our goal is to show that this
is either a long range or power-law correlation for arbitrary
$\beta$. In other words, Eq.~\ref{tising} is less likely to
disorder than the ordinary 2d Ising model. This result can be
naively understood as follows: the ordinary 2d Ising model is
disordered at high temperature (small $\beta$) due to the
proliferation of Ising domain walls. But in the current model, due
to the $(-1)$ sign associated with each domain wall, the
proliferation of domain walls is suppressed, thus eventually the
current Ising model \eqnref{tising} is not completely disordered
even for small $\beta$.

This Ising model is dual to a loop model with the following
partition function: \beqn\label{eq: O(n) model} Z =
\sum_{\mathcal{C}} K^L n^{N_d}, \eeqn where loops are the domain
walls of the original Ising model, $K = \exp( - 2\beta)$ is the
loop tension, $n = -1$ is the loop fugacity, $L$ is the total
length of loops, and $N_d$ is the total number of closed loops. If
the loops do not cross, then according to
Ref.\,\cite{Nienhuis:1982la}, by tuning $K$ there is a phase
transition between a small loop phase (which corresponds to the
Ising ordered phase) for small $K$, and a dense loop phase for
large $K$. The critical point and the dense loop phase are both
critical with power-law correlations, and they correspond to two
different conformal field theories with central charges $c = -
3/5$ and $c = -7$ respectively. If the loops are allowed to cross,
the dense loop phase is driven to a different conformal field
theory with $c = -2$, which is described by free symplectic
fermions.\cite{Jacobsen:2003pb}

The Ising order parameter $\sigma_i$ corresponds to the ``twist"
operator of the loop model, because $\sigma_i$ changes its sign
when it crosses a loop. The twist operator is well-studied at the
critical point of loop models, and in our case with $n = -1$, at
the critical point between small and dense loop phases the scaling
dimension of the twist operator is $-1/10$~\cite{cardyloop}, which
is confirmed by our numerical calculation.

The tensor renormalization group (TRG) method\cite{Levin:2007it,
Gu:2009eu} has been applied to loop models in
Ref.\,\cite{Gu:2009dn}. Here we use the same approach to study the
twist operator correlations for the loop model in \eqnref{eq: O(n)
model}. For simplicity we forbid the loops to cross, so the model
never develops antiferromagnetic order even for negative $\beta$.
For positive large $\beta$, the strange correlator is long-ranged,
see \figref{fig: LG}(a). As $\beta$ decreases, the correlator
grows and diverges at the critical point $\beta_c\simeq 0.521$
with a power-law $C(r, r^\prime) \sim |r - r^\prime|^{0.199}$ as
shown in \figref{fig: LG}(b), which confirms the theoretical
prediction of scaling dimension $-1/10$ of twist
operator~\cite{cardyloop}. Theoretically the entire dense loop
phase (when $ \beta < \beta_c$) should be controlled by one stable
conformal field theory fixed point. Our numerical results suggest
that this fixed point is likely around $\beta \sim -0.1816$, the
power-law behavior of $C(r, r^\prime)$ at this point
(Fig.~\ref{fig: LG}) is qualitatively consistent with the
conclusion of this paper. \footnote{For $\beta$ far away from this
fixed point, the finite system size and error bar, as well as the
incommensurate oscillation of the strange correlator make it more
difficult to extract a conclusive scaling dimension of $\sigma_i$.
But we expect $C(r, r^\prime)$ to crossover back to the same
scaling behavior as the stable fixed point $\beta \sim -0.1816$ in
the infrared limit for arbitrary $\beta < \beta_c$. Our result may
have also been strongly affected by our choice of microscopic
rules for loops close to each other. More recent studies by
Scaffidi and Ringel~\cite{zohar} on the Levin-Gu model on a
triangular lattice have successfully extracted a scaling dimension
consistent with the Coulomb gas prediction of the dense loop
phase~\cite{cardyloop}. }

\begin{figure} %[htbp]
\begin{center}
\includegraphics[width=200pt]{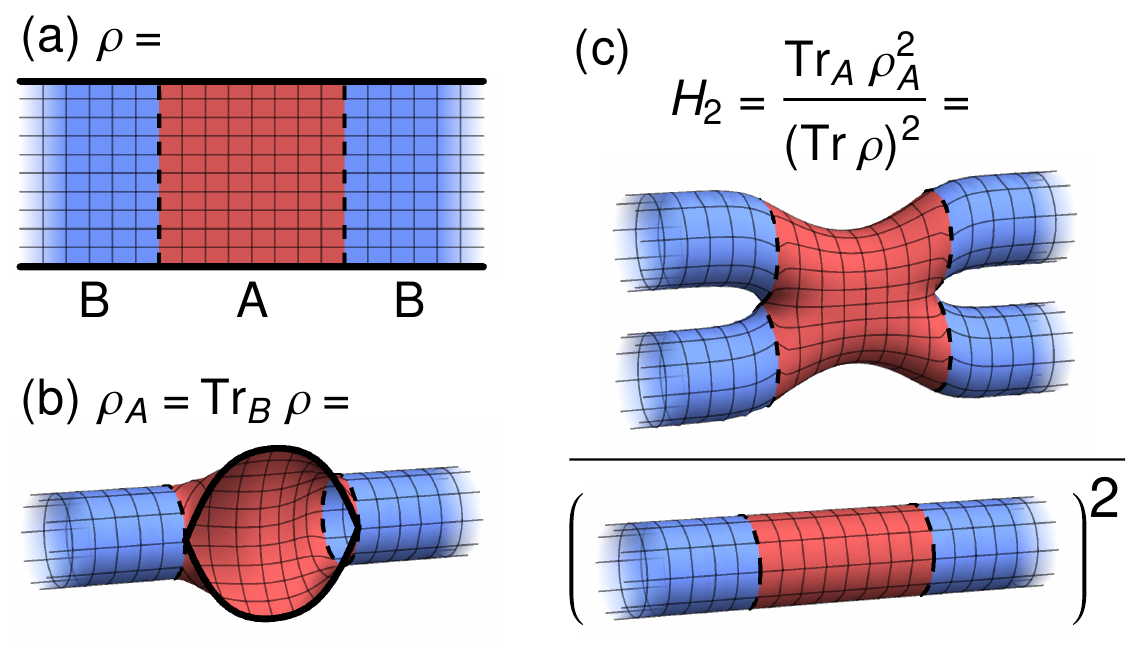}
\caption{(color online). (a) Under Lorentz transformation, the
density matrix of the SRE edge states is mapped to the overlap
between bulk ground state wave functions on a manifold with open
boundaries in one direction. The edge manifold may be partitioned
into the regions $A$ (red) and $B$ (blue). (b) The reduced density
matrix in the region $A$ of the edge states corresponds to joining
the boundaries of $B$ together. (c) $\text{Tr}\rho_A^2$ is given
by doubling $\rho_A$ and sealing the boundaries of regions $A$
with each other, resulting in the pants (double torus) topology.
$\text{Tr}\rho$ is simply obtained by rolling up (a). Their ratio
gives the R\'enyi entropy $H_2$.} \label{fig: pants}
\end{center}
\end{figure}

We have checked that the ordinary free electron $3d$ topological
insulator also gives us a very clear power-law decay of strange
correlator. However, in general a strongly interacting SRE state
in three dimensional space can be more complicated, because its
two dimensional edge can be (1) a gapless $(2+1)d$ conformal field
theory, (2) long range order that spontaneously breaks symmetry,
(3) two dimensional topological phase~\cite{vishwanathsenthil}.
Based on our Lorentz transformation argument, it is possible that
$\langle \Omega | \Psi \rangle$ is mapped to the partition
function of a topological phase, then in this case the strange
correlator $C(r, r^\prime)$ may also be short ranged. Thus for 3d
SRE states, besides the strange correlator, we also need another
method that diagnoses the situation when $\langle \Omega | \Psi
\rangle$ corresponds to a topological phase partition function.

The method we propose is illustrated in Fig.~\ref{fig: pants},
where the horizontal direction represents the XY plane, while the
vertical direction is the $z$ axis of the three dimensional space.
We can first calculate the overlap between the given 3d wave
function $|\Psi\rangle$ and the trivial wave function on a 3d
``pants"-like manifold in Fig.~\ref{fig: pants}$c$ ($\langle
\Omega | \Psi\rangle_{\mathrm{pants}}$), which after Lorentz
transformation becomes $\text{Tr}\rho_A^2$ at the edge, where
$\rho_A$ is the reduced density matrix of subsystem $A$ at the
boundary. The following quantity after Lorentz transformation
becomes the R\'{e}nyi entanglement entropy of the edge topological
phase: \beqn S = - \log\left( \frac{\langle \Omega |
\Psi\rangle_{\mathrm{pants}}}{(\langle \Omega |
\Psi\rangle_{\mathrm{cylinder}})^2} \right). \eeqn This quantity
should scale as $S = \alpha L - \gamma$, where $\gamma$ is the
analogue of the topological entanglement entropy of the edge
topological phase~\cite{wenentropy,kitaeventropy}. Thus a $3d$
wave function $|\Psi\rangle$ is still a nontrivial SRE state as
long as $\gamma$ defined above is nonzero, even if this wave
function has a short range strange correlator. We will leave the
detailed study of this proposal to future work.

In summary, we have proposed a general method to diagnose $1d$ and
$2d$ SRE states based on their bulk ground state wave functions.
We expect our method to be useful for future numerical studies of
SRE states. In Ref.~\cite{wang1,wang2,wang3,wang4}, it was
proposed that interacting fermionic topological insulators and
topological superconductors can be characterized by the full
fermion Green's function; Ref.~\cite{zaletel} proposed a method to
diagnose bosonic SPT states characterized by group cohomology. The
method proposed in our current paper is applicable to both
fermionic and bosonic SRE states.

The authors are supported by the the David and Lucile Packard
Foundation and NSF Grant No. DMR-1151208. The authors also thank
John Cardy, Adam Nahum, Jesper Jacobsen, T. Senthil, T. Scaffidi
and Z. Ringel for very helpful discussions.

\bibliography{temp}

\end{document}